\documentclass[showpacs,showkeys,preprint]{revtex4}
\usepackage{amsmath}
\usepackage{graphics}
\begin{document}

\title{Minimal speed of fronts of  reaction-convection-diffusion equations}
\author{R. D. Benguria}
\affiliation{
 Facultad de F\'\i sica\\
    Pontificia Universidad Cat\'olica de Chile\\
           Casilla 306, Santiago 22, Chile}
\author{ M. C. Depassier}
\affiliation{
 Facultad de F\'\i sica\\
    Pontificia Universidad Cat\'olica de Chile\\
           Casilla 306, Santiago 22, Chile}
\author{V. M\'endez}
\affiliation{Facultat de Ciencies de la Salut\\ Universidad
Internacional de Catalunya \\ Gomera s/n 08190 Sant Cugat del
Valles, Barcelona, Espa\~na.}

\date{\today}

\begin{abstract}
We study the minimal speed of propagating fronts of  convection
reaction diffusion equations of the form  $u_t + \mu\, \phi(u) u_x
= u_{xx} +f(u)$ for positive reaction terms with $f'(0 >0$. The
function $\phi(u)$ is continuous and vanishes at $u=0$.  A
variational principle for the  minimal speed of the waves is
constructed from which upper and lower bounds are obtained.  This
permits the a priori assesment of the effect of the convective
term on the minimal speed of the traveling fronts. If the
convective term is not strong enough, it produces no effect on the
minimal speed of the fronts. We show that if $f''(u)/\sqrt{f'(0)}
+ \mu \phi'(u) < 0$, then the minimal speed is given by the linear
value $2 \sqrt{f'(0)}$, and the convective term has no effect on
the minimal speed. The results are illustrated by applying them to
the exactly solvable case $u_t + \mu u u_x = u_{xx} + u (1 -u)$.
Results are also given for the density dependent diffusion case
$u_t + \mu\, \phi(u) u_x = (D(u)u_x)_x +f(u)$.
\end{abstract}

\pacs{ 87.23.Cc, 04.20.Fy,  05.60.Cd,  02.60.Lj}
\keywords{Reaction-diffusion-convection, population dynamics,
variational principles}

\maketitle

\section{Introduction}
\label{intro} The reaction diffusion equation $u_t = u_{xx} +
f(u)$ has been employed as a simple model of phenomena in
different areas, population growth, chemical reactions, flame
propagation and others.  In the classical Fisher case
\cite{Fisher}, $ f(u) = u (1-u)$, a front propagating with speed
$c_{\mbox{kpp}}=2$ joins the two equilibrium points \cite{KPP}.
The time evolution for general reaction terms was solved by
Aronson and Weinberger [AW] \cite{AW78} who showed that
sufficiently localized initial conditions evolve into a front
which propagates with speed $c_*$ such that $ 2 \sqrt{f'(0)} \le
c_* \le 2 \sqrt{\sup(f(u)/u)}$. The asymptotic speed of
propagation is the minimal speed for which a monotonic front
joining the stable to unstable equilibrium point exists. Existence
proofs  give limited  quantitative  information on the dependence
of the speed of the front on the parameters of the problem
\cite{Heinze}. For this reason different variational methods have
been developed. For the one dimensional case it has been shown
that this minimal speed can be derived  either from a local
variational principle of the minimax type \cite{HR75}, or from an
integral variational principle \cite{BD96a,BD96c}. Minimax
variational principles  for the speed of fronts in several
dimensions and for inhomogeneous environments have also been
established \cite{Heinze,Volpert}.

In many processes, in addition to diffusion, motion can also be
due to advection or convection. Nonlinear advection terms arise
naturally, for example, in the motion of chemotactic cells. In a
simple one dimensional model,  denoting by $\rho$ the density of
bacteria, chemotactic to a single chemical element of
concentration $s(x,t)$ the density evolves according to
$$
\rho_t  = [ D \rho_x - \rho \chi s_x]_x +  f(\rho),
$$
where diffusion, chemotaxis and growth have been considered. There
is some evidence \cite{redbook} that, in certain cases,  the rate
of chemical consumption is due mainly  to the ability of the
bacteria to consume it. In that case
$$
s_t = - k \rho,
$$
where diffusion of the chemical has been neglected (arguments to
justify this approximation, together with the choice of constant
$D$ and $\chi$ are given in \cite{redbook}). If we now look for
traveling wave solutions $s= s(x - c t)$, $\rho = \rho(x - c t)$,
then $s_t = - c s_x$, therefore $s_x = k \rho /c$, and the problem
reduces to a single differential equation for $\rho$, namely,
\begin{equation}
\rho_t  =  D \rho_{xx} - \frac{\chi k}{c} (\rho^2)_x +  f(\rho).
\label{eq:Segel}
\end{equation}
The more elaborate models of Keller and Segel for chemotaxis \cite{KellerSegel},
 which include diffusion of the chemical and other effects, have been considered to
 explain chemotactic collapse (\cite{Childress,Herrero} and references therein) and
 other phenomena.  The derivation of these equations from transport theory and the
 assumptions involved in them have been studied recently \cite{Othmer}.
In addition to these biological processes, equations analogous to
(\ref{eq:Segel}) appear when modeling the Gunn effect in
semiconductors and in other physical phenomena
\cite{Lika,Bonilla}. Equation (\ref{eq:Segel}) for a Fisher type
reaction term $f(u) = u(1-u)$ has been studied in \cite{Murray2}.
An extensive study of the existence of traveling waves of
nonlinear diffusion-reaction-convection equations which includes a
review of many results to which we refer for additional references
is contained in \cite{Gilding}.

In this work we concentrate on the  equation with a general
convective term which, suitably scaled, we write as
\begin{equation}
\label{eq:pde} u_t + \mu\, \phi(u)\, u_x = u_{xx} + f(u),
\end{equation}
where the reaction term $f(u)$ is a continuous function with
continuous derivative in $[0,1]$ and satisfies
$$
f(0) = f(1)= 0, \qquad f'(0) >0 , \qquad f'(1) <0 \qquad {\rm and}
\qquad f > 0\quad  {\rm in}\quad (0,1) .
$$
The function $\phi(u)$ is a continuous function with continuous
derivative in $[0,1]$.  Without loss of generality we may assume
that in addition $\phi(0) = 0$, since otherwise only a uniform
shift in the speed is introduced. The parameter $\mu$ is positive.

For equation (2), the existence of monotonic decaying traveling
fronts $u(x- c t)$ for any wave speed greater than a critical
value $c_*$ has been proved recently \cite{Malaguti}. Moreover, in
\cite{Malaguti} the  following estimate for the threshold value
$c_*$ is obtained,
\begin{equation}
2\sqrt{f'(0)} \le c_* \le 2 \sqrt{\sup_{u\in(0,1]} \frac{f(u)}{u}}
+ \max_{u\in[0,1]}\mu \phi(u). \label{Malaguti}
\end{equation}
Analogous results for density dependent diffusion are also
established in \cite{Malaguti}. The convergence of some initial
conditions to a monotonic traveling front has been proved
\cite{Crooks2003} for systems in which the minimal speed is
strictly greater than the linear value $c_L = 2\sqrt{f'(0)}$.

We will show that the minimal speed $c_*$ for the existence of a
monotonic decaying front $u(x-c t)$ joining the stable equilibrium
$u=1$ to the unstable equilibrium $u=0$ obeys the variational
principle
\begin{equation}
c_* = \sup_{g\in {\cal S}} {\cal E}(g) , \label{eq:var}
\end{equation}
with
\begin{equation}
{\cal E}(g) = \frac{\int_0^1 (2 \sqrt{f(u) g(u) (- g'(u))} + \mu
\phi(u) g(u) )\,d u} { \int_0^1 g(u)\, d u} \label{ecal}
\end{equation}
and the supremum is taken over the set ${\cal S}$ of all positive,
monotonic decreasing functions $g(u)$ for which the integrals in
(\ref{ecal}) exist and $g(1)=0$.
 From here it will follow that
\begin{equation}
2\sqrt{f'(0)}  \equiv c_L \le c_*\le  \inf_{\alpha >0} \sup_{u\in
[0,1]} (\alpha + \frac{1}{\alpha} f'(u) + \mu \phi(u)).
\label{eq:ooso}
\end{equation}
From the variational expression (\ref{eq:var}) one may obtain the
value of the minimal speed with any desired accuracy, and the
inequalities (\ref{eq:ooso}) enable us to characterize the
reaction terms for which the speed is the linear  value $c_L$.
More precisely, if
$$
\frac{f''(u)}{\sqrt{f'(0)}} + \mu \phi'(u) < 0, \qquad \mbox{then}
\quad c_* = 2 \sqrt{f'(0)}.
$$

The bound (\ref{Malaguti}) is also derived from the variational
principle. The generalization to density dependent diffusion is
given as a direct extension of the previous results.

\section{Minimal speed of traveling fronts}
\label{sec1}

Traveling monotonic decaying  fronts  $u(x-ct)$ of (\ref{eq:pde})
satisfy the ordinary differential equation
$$
u_{zz} + (c - \mu \phi(u) )  u_z + f(u) = 0 \qquad \lim_{z
\rightarrow -\infty }u = 1, \quad \lim_{z \rightarrow \infty }u =
0,  \quad u_z < 0,
$$
where $z = x -c t$. It is convenient to work in phase space;
defining as usual $p(u) = - u_z$, the problem reduces to finding
the solutions of
\begin{equation}
p(u)\, \frac{d p(u) }{du} - (c - \mu\, \phi(u))\, p(u) + f(u) = 0
, \label{phase}
\end{equation}
with
$$
p(0) = p(1) = 0 , \qquad {\rm and}\,\, p> 0.
$$

We first perform the linear analysis around the endpoints $u=0$
and $u=1$ which may provide restrictions on the allowable speed.
These results will also be needed when proving the existence of a
variational principle. Near $u=0$,  $p(u) = m u$ where $m$ is the
larger root of $m^2 -F(0) m + f'(0) = 0. $ For convenience, we
have defined $F(u) = c - \mu \phi(u)$. This root is given by
\begin{equation*}
m = \frac{ F(0) + \sqrt{F(0)^2 - 4 f'(0)}}{2}.
\end{equation*}
The condition that $m$ be real imposes the restriction $F^2(0) \ge
4 f'(0)$.  Written explicitly this bound is,
\begin{equation}
c \ge 2 \sqrt{f'(0)}  \equiv c_L. \label{csubl}
\end{equation}
Near $u=1$,  $p = r (1-u)$ where $r$ is the  positive root of $
r^2 + F(1) r + f'(1) =0, $ namely,
\begin{equation}
r = \frac{ - F(1) + \sqrt{ F^2(1) - 4 f'(1)}}{2}. \label{r}
\end{equation}
No additional restriction on the range of allowable speeds is
imposed from the expression above, since by hypothesis $f'(1) <
0$.

In addition to the linear constraint (\ref{csubl}), a simple
constraint is found from direct integration of  Eq.(\ref{phase}).
Dividing by $p(u)$ and integrating between 0 and 1, we have
$$
c = \mu \int_0^1 \phi(u) d\,u + \int_0^1 \frac{f(u)}{p(u)}d\, u.
$$
Since $f$ and $p$ are positive in $(0,1)$, we obtain
\begin{equation*}
c >  \mu \int_0^1 \phi(u) d\,u.
\end{equation*}

\subsection{Variational principle}
\label{sec2}

In this section we construct a variational principle from which
the exact speed of the front may be calculated.  Let $g$ be any
positive function in (0,1) such that $h = - dg/du > 0$.
Multiplying Eq.(\ref{phase})) by $g/p$ and integrating with
respect to $u$ we find that
$$
c \int_0^1 g\, du = \int_0^1 \left( h\, p + \frac{f}{ p}\, g
\right) du + \mu \int_0^1 \phi g du $$ where the first term on the
right-hand side is obtained after integration by parts. However
since  $p,\,h,\, f, {\rm and}\, g$ are positive, we have that for
every fixed $u$
$$
h\,p + \frac{f\, g}{ p} \ge 2 \, \sqrt{f\, g\, h}
$$
so that,
\begin{equation}
c\, \ge \frac {\int_0^1 [ 2 \sqrt{ f\, g\, h} + \mu  \phi g ] d u
}{ \int_0^1
 g\, du} = {\cal E}(g)
\label{tut}
\end{equation}
 To show that this is a variational principle we must prove that there exists a
 function $g = \hat g$ for which equality holds.
Equality is attained for $g = \hat g$ such that
$$
p h = - p\hat g'
 = \frac{\hat g f}{ p}.
$$
Using Eq.(\ref{phase}) to eliminate $f(u)$ we have that
$$
\frac{\hat g'(u)}{\hat g(u)} - \frac{p'(u)}{p(u)} =  -
\frac{F(u)}{p(u)},
$$
which can be integrated to obtain
\begin{equation}
\hat g(u) = p(u) \exp \left[ -\int_{u_0}^u \left( \frac{F(t)}{
p(t)} \right) dt \right], \qquad \text{with}\quad  0 < u_0 < 1.
\label{g}
\end{equation}
Since $p$ vanishes at 0 and 1, we must analyze the behavior of
$\hat g$ at these points in order to ensure the convergence of the
integrals in (\ref{tut}).

At $u=1$, since $p(u) = r (1-u) $, and since $F(u)$ is continuous
at 1,
 we obtain that
\begin{equation*}
\hat g \sim r (1 - u)^{(1 + F(1)/r)}.
\end{equation*}
From the expression for $r$, (\ref{r}) we see, that, since $f'(1)
< 0$, for any value of $F(1)$ the exponent $1 + F(1)/r$ is
positive, hence $\hat g(1) =0$.

Near $u=0$, since $p = m u$ and since $F$ is continuous at zero,
we find that
\begin{equation*}
\hat g  \sim \frac{m}{u^{(F(0)/m) - 1}}.
\end{equation*}
The integrals in (\ref{tut}) converge provided
$$
\frac{F(0)}{m} - 1 < 1.
$$
This condition is satisfied whenever $c_* > c_L$. That is,
whenever $c_*> c_L$ there exists a function $g = \hat g$ for which
equality holds in (\ref{tut}) or , equivalently, $c_* = \max {\cal
E}(g) = {\cal E}(\hat g)$.

On the other hand, when $c_* = c_L$, there does not exist a
function $\hat g$ in the set ${\cal S}$ of admissible functions
for which equality holds in (\ref{tut}). Consider however the
trial functions
$$
g_\alpha(u) = u^{\alpha-1} -1.
$$
Clearly $g_\alpha \in {\cal S}$ for any $0<\alpha<1$. Moreover one
can check that (see the Appendix)
$$
\lim_{\alpha\rightarrow 0} {\cal E}(g_\alpha) = c_L = c_*,
$$
therefore
$$
c_* = \sup_{g\in {\cal S}} {\cal E}(g)
$$
 in this case.
Notice that in this last case the maximum is not attained since
the limiting function $g_0$ does not belong to ${\cal S}$. This
concludes the proof of our variational principle.

\subsection{Upper and Lower Bounds}
\label{sec3}

 The variational principle provides lower bounds with suitably chosen trial
 functions, which can be arbitrarily close to the exact value of the speed.
  The fact that it is a variational principle for which equality holds, enables
  one to obtain also an upper bound to the speed.

To obtain an upper bound we use the fact that
\begin{equation}
2 a b \le \alpha a^2 + \frac{1}{\alpha} b^2, \qquad
\text{with}\quad  \alpha>0. \label{ineq}
\end{equation}
Then the following inequality holds:
$$
2 \sqrt{fgh}= 2 g \sqrt{fh/g} \le g \left(\alpha  +
\frac{1}{\alpha}\frac{fh}{g}\right),
$$
where we used the inequality above with $a=1, b= \sqrt{fh/g}$.

Then,
$$
c_* = \sup_g \left[ \frac{\int_0^1 ( 2 \sqrt{ f\, g\, h}+ \mu
\phi g ) du }{\int_0^1
 g\, du}\right]
\le \sup_g \frac {{\int_0^1 g \left[ \alpha + f h/(\alpha g)+ \mu
\phi \right] du}}{{\int_0^1
 g\, du}}.
$$
The second term in the last expression can be  integrated by
parts.  The boundary term $f g|_0^1$ vanishes and we  obtain
\begin{eqnarray}
c_* &\le& \sup_g  \frac{{\int_0^1 g ( \alpha   + f'/\alpha + \mu
\phi) du}}{{\int_0^1
 g\, du}} \nonumber \\
&\le& \sup_{u \in [0,1]} \left[\alpha + \mu \phi +
\frac{1}{\alpha} f'\right] \label{aaa}.
\end{eqnarray}

The above inequality holds for any positive $\alpha$, hence
\begin{equation}
c_*\le  \inf_\alpha \sup_{u \in [0,1]} \left[\alpha +
\frac{1}{\alpha} f'(u) + \mu \,\phi(u)\right]. \label{upper}
\end{equation}

The bound (\ref{upper})  in the
 case $\mu =0$ differs from the classical Aronson-Weinberger \cite{AW78} result
 for fronts of the parabolic reaction diffusion equation
$c \le  c_{AW} = 2 \sup_{u\in [0,1]} \sqrt{f(u)/u}$. The bound
(\ref{Malaguti}) obtained in \cite{Malaguti} on the other hand,
reduces, when $\mu=0$ to the classical Aronson-Weinberger result.
Here we show that this last bound can be derived from the
variational principle as well.   Using the inequality
(\ref{ineq}), now with $a = \sqrt{f g/u}$ and $b = \sqrt{ u h}$,
we have that
$$
2\int_0^1 \sqrt{f g h} d u \le \int_0^1 \left[ \alpha \frac{f
g}{u} + \frac{1}{\alpha} h u\right] d u = \int_0^1 \left[ \alpha
\frac{f g}{u} + \frac{1}{\alpha} g \right] d u,
$$
where the last expression is obtained after integrating the second
term by parts. We have then
\begin{equation*}
c_* \le \sup_g  \frac{{\int_0^1 g ( \alpha \frac{f}{u}  +
\frac{1}{\alpha} + \mu \phi) du}}{{\int_0^1
 g\, du}} \le \sup_{u\in [0,1]} \left[\alpha \frac{f}{u}
  + \frac{1}{\alpha} + \mu \phi \right].
\end{equation*}
Choosing $\alpha = 1/ \sup \sqrt{f/u}$ we obtain
$$
c_* \le \sup_{u\in ]0,1]} \left[ 2 \sqrt{\frac{f(u)}{u}}\right] +
\mu \max_{u\in [0,1]}\phi(u) .
$$

In the classical AW case $\mu=0$, we know that when the reaction
term is concave then $\sup 2 \sqrt{f(u)/u} = 2 \sqrt{f'(0)}$. In
this case the upper bound coincides with the linear lower bound
$c_L$ and the minimal speed is univocally determined. A similar
criterion can be obtained in the present problem. The minimal
speed for the existence of a front  is known unambiguosly to be
the linear value whenever the upper bound (\ref{upper}) coincides
with the linear lower bound $c_L$. A sufficient condition for this
to occur is that the supremum of the function $ K(u) =   \alpha +
f'(u)/\alpha + \mu \phi(u)$ in $(0,1]$ does not exceed the value
of $K$ at the origin. Effectively, if  $\sup_u K(u) = K(0) =
\alpha +  f'(0)/\alpha$; minimizing with respect to $\alpha$ we
obtain $\alpha = \sqrt{f'(0)}$ and the upper bound is precisely
the linear value. A sufficient condition that guarantees that the
maximum (supremum) of $K$ occurs at zero is that $K(u)$ be
decreasing. With $\alpha = \sqrt{f'(0)}$ this condition is
\begin{equation*}
\frac{f''(u)}{\sqrt{f'(0)}} + \mu \phi'(u) < 0.
\end{equation*}
Whenever this condition is fulfilled, for all $u$, we know that
the minimal speed of a monotonic front is the linear value $c_L =
2 \sqrt{f'(0)}$. Again as it occurs in the standard case $\mu=0$,
this condition is sufficient but not necessary.

\section{An exactly solvable case}

Here we illustrate the above results by applying them to the
exactly solvable case discussed in  \cite{Murray2}: a Fisher type
reaction term $f(u) = u(1-u)$ and the simplest convective term
$\phi(u) = u$. By means of a phase space analysis Murray found
\cite{Murray2} that the minimal speed for the existence of a
monotonic decaying front is
\begin{equation}
c_* = \left\{ \begin{array}{ll}
              \frac{2}{ \mu} + \frac{\mu}{ 2}
             & \mbox{if $\mu > 2$} \\
             2  &\mbox{if $\mu \le 2$}
           \end{array} \right.
\label{Murray}
\end{equation}
Here we show that the results of the previous section allow the
exact determination of the speed.

In this example the linear marginal stability value is given by
$c_L = 2$. We first use the variational principle to obtain a
lower bound. Take the trial function
$$
g(u) = \left(\frac {1-u}{ u} \right)^\lambda \qquad
\mbox{with}\quad 0 < \lambda < 1.
$$
A straightforward integration of  (\ref{tut}) leads to
$$
c \ge 2 \sqrt{\lambda} + \frac{\mu}{ 2} (1 - \lambda) \equiv
c(\lambda).
$$
If $\mu > 2$ the maximum of $c(\lambda)$ occurs for $\lambda =
4/\mu^2$ and it is given by $2/\mu + \mu/2$. For $\mu \le 2$,
however, the supremum of $c(\lambda)$ occurs as $\lambda \to 1$.
We have then
$$
c_* \ge \sup c(\lambda) = \frac{2}{ \mu} + \frac{\mu}{ 2} \qquad
\mbox{for} \quad\mu > 2,
$$
and
$$
c_* \ge \sup c(\lambda) = 2 \qquad \mbox{for}\quad\mu \le 2.
$$

To obtain an upper bound we use (\ref{aaa}), that is,
$$
c_* \le \sup_{u\in [0,1]} ( \alpha + \frac{1}{\alpha} + u ( \mu -
\frac{2}{\alpha}) ) \quad \forall \alpha > 0.
$$
We will separate the two cases, $\mu \le 2$ and $\mu > 2$.

If $\mu \le 2$, choose $\alpha = 1$, then
$$
c_* \le \sup_{u\in [0,1]} ( 2 + u (\mu-2) ) =2.
$$
If $\mu > 2$ choose $\alpha = 2/\mu$, then
$$
c_* \le \sup_{u\in [0,1]} \left( \frac{2}{\mu} + \frac{\mu}{2
}\right) =  \frac{2}{\mu} + \frac{\mu}{2 }.
$$

The lower bound obtained from the variational expression coincides
with the upper bound obtained from (\ref{upper}), therefore we
know with certainty that the minimal speed is indeed,
(\ref{Murray}) which had been previously demonstrated by phase
space methods.

Notice that (\ref{Malaguti}) constitutes a poorer bound in this
case. Effectively, from (\ref{Malaguti}) it follows that $c_* \le
2 +\mu$.

\section{Density dependent diffusion}

The effect of the convective term on the minimal speed of fronts
of the reaction diffusion equation for non constant difffusion
follows in a simple way from the previous results. Consider
traveling fronts of the equation
\begin{equation*}
u_t + \mu\, \phi(u)\, u_x = (D(u)u_x)_x + f(u),
\end{equation*}
where $f(u)$ and $\phi(u)$ satisfy the properties spelled in the
previous sections. The diffusion coefficient $D(u)$ is continuous
and $D(u) >0$ in $(0,1]$.  $D(0)$ is either positive or zero. By a
suitable change of variables \cite{Hadeler2,Engler} the equation
for the fronts is reduced of the usual reaction diffusion equation
with a reaction term $\tilde f(u) = D(u) f(u)$.
 This reaction term satisfies $\tilde f >0$, and $\tilde f'(0) = D(0) f'(0)$.
We must distinguish two cases. If $D(0) \neq 0$, then $\tilde f$
satisfies the same properties as $f$ and the results of the
previous sections can be applied directly. If $D(0) =0$, then
$\tilde f'(0) = 0$ and we expect a sharp wavefront. In this case
it has been shown \cite{AW78} that the front approaches $u=0$ as
$c u$. A variational principle exists also in this case
\cite{BDdensity}. We have then that, in both cases, the minimal
speed of the wavefronts is given by
$$
c_* = \sup_g \left[ \frac{\int_0^1 (2 \sqrt{D(u) f(u) g(u) (-
g'(u))}+ \mu  \phi(u) g(u))\, d u} { \int_0^1 g(u)\, du} \right]
$$
where the supremum is taken over all positive monotonic decreasing
functions $g(u)$ for which the integrals exist and $g(1) =0$.
Upper and lower bounds can be obtained following the methods of
the previous sections.
 We do not spell them out here.

\section{Summary}

We have studied the effect of a convective term on the speed of
monotonic reaction-diffusion fronts.  The minimal speed for the
existence of fronts has been showed to derive from a variational
principle from which the exact speed can be determined in
principle. The existence of this variational characterization
permits the obtention of upper and lower bounds. The classical
result that establishes that for concave reaction terms, the
minimal speed of the fronts  is the linear or KPP value is
extended to the case where convective terms are present. The
extension to the the case of density dependent diffusion has been
given for positive diffusion terms.

We have found that a convective term increases the minimal speed
of the traveling front only if it is sufficiently strong, if not,
the minimal speed is determined by the reaction term alone.

\section*{Acknowledgements}
R. D. Benguria and M. C. Depassier  acknowledge partial support
from Fondecyt under grants 1020844 and 1020851. V. M\'endez
acknowledges support from grants BFM2000-0351 and SGR-2001-00186.

\appendix
\section{}

In this appendix we prove that
$$
\lim_{\alpha\rightarrow 0} {\cal E}(g_\alpha) = c_L
=2\sqrt{f'(0)},
$$
where $g_\alpha= u^{\alpha-1}-1$ with $0 < \alpha < 1$.

Since $\int_0^1 g_{\alpha}(u) d u = (1-\alpha)/\alpha$, and
$h_\alpha= (1-\alpha) u^{\alpha-2}$, we may write
$$
{\cal E}(g_\alpha ) = J_1(\alpha) + J_2(\alpha),
$$
where
$$
J_1(\alpha) = \frac{2 \alpha}{\sqrt{1-\alpha}} \int_0^1 \sqrt{
f(u) (u^{2\alpha-3} - u^{\alpha-2})}\, d u,
$$ and
$$
J_2(\alpha) = \frac{\mu \alpha}{1-\alpha}\int_0^1 \phi(u)
(u^{\alpha-1} -1) d u.
$$
Since $\phi(0)=0$ and $\phi$ is continuous the integral in $J_2$
has a finite value when $\alpha =0$. Then, due to the overall
multiplicative factor of $\alpha$, we see that
$$
\lim_{\alpha \rightarrow 0} J_2(\alpha) = 0.
$$
To show that $\lim J_1 (\alpha) = 2 \sqrt{f'(0)}$, as $\alpha
\rightarrow 0$, write
$$
J_1(\alpha) = \frac{2 \alpha}{\sqrt{1-\alpha}}\int_0^1 \sqrt{ u
f'(0) (u^{2\alpha-3} - u^{\alpha-2})} \,d u + K(\alpha),
$$
where $$ K(\alpha) = \frac{2 \alpha}{\sqrt{1-\alpha}}\left[
\int_0^1 \sqrt{ f(u) (u^{2\alpha-3} - u^{\alpha-2})} d u -
\int_0^1 \sqrt{ u f'(0) (u^{2\alpha-3} - u^{\alpha-2})} d
u\right].
$$
The first integral is
$$
\frac{2 \alpha}{\sqrt{1-\alpha}}\int_0^1 \sqrt{ u f'(0)
(u^{2\alpha-3} - u^{\alpha-2})} d u = \frac{\sqrt{ \pi
f'(0)}}{\sqrt{1-\alpha}} \frac { \Gamma( \frac{1}{1-a})}
{\Gamma(\frac{-3+a}{2(a-1)})}.
$$
 Now we prove that $\lim_{\alpha\rightarrow 0} K(\alpha) = 0$.
 $$
 |K(\alpha)| \le  \frac{2 \alpha}{\sqrt{1-\alpha}}\int_0^1 | \sqrt{ f(u)
  (u^{2\alpha-3} - u^{\alpha-2})} -  \sqrt{ u f'(0) (u^{2\alpha-3} - u^{\alpha-2})}\, | d u.
 $$
 But $| \sqrt{a} - \sqrt{b}| \le \sqrt{|b - a|}$, therefore
 $$
 | K(\alpha)| \le \frac{2 \alpha}{\sqrt{1-\alpha}}\int_0^1 \sqrt{ |f(u) - u f'(0)|
 (u^{2\alpha-3} - u^{\alpha-2})} d u.
 $$
 Since $f(u)$ and its derivative are continuous, in $[0,1]$, there exist $d>0$,
 $q>0$ such that
$$
\frac{ | f(u) - u f'(0)|}{u} < d u^q.
$$
In particular, if $f(u)$ is analytic in a neighborhood of 0,
$q=1$. Using this inequality in the expression above, we have that
$$
|K(\alpha)| \le \frac{2 \alpha}{\sqrt{1-\alpha}}\int_0^1 \sqrt{ d
u^{q+1} (u^{2\alpha-3} - u^{\alpha-2})} d u.
$$
Finally, since $u^{2\alpha-3} - u^{\alpha-2} < u^{2\alpha-3}$,
$$
 |K(\alpha)| \le \frac{2 \alpha}{\sqrt{1-\alpha}} \sqrt{d} \int_0^1
 u^{\alpha-1 + q/2} du =
 \frac{2 \alpha}{\sqrt{1-\alpha}} \frac{\sqrt{d}}{\alpha + q/2}.
 $$
 Therefore, $\lim_{\alpha\rightarrow 0} |K(\alpha)| =0$.

  To sum up,
 $$
 \lim_{\alpha\rightarrow 0}{\cal E}(g_\alpha ) = \lim_{\alpha\rightarrow 0}
 \frac{\sqrt{ \pi f'(0)}}{\sqrt{1-\alpha}}
\frac { \Gamma( \frac{1}{1-a})} {\Gamma(\frac{-3+a}{2(a-1)})}  = 2
\sqrt{f'(0)}.
$$

\end{document}